
\magnification=\magstep0
\font \authfont               = cmr10 scaled\magstep4
\font \fivesans               = cmss10 at 5pt
\font \headfont               = cmbx12 scaled\magstep4
\font \markfont               = cmr10 scaled\magstep1
\font \ninebf                 = cmbx9
\font \ninei                  = cmmi9
\font \nineit                 = cmti9
\font \ninerm                 = cmr9
\font \ninesans               = cmss10 at 9pt
\font \ninesl                 = cmsl9
\font \ninesy                 = cmsy9
\font \ninett                 = cmtt9
\font \sevensans              = cmss10 at 7pt
\font \sixbf                  = cmbx6
\font \sixi                   = cmmi6
\font \sixrm                  = cmr6
\font \sixsans                = cmss10 at 6pt
\font \sixsy                  = cmsy6
\font \smallescriptfont       = cmr5 at 7pt
\font \smallescriptscriptfont = cmr5
\font \smalletextfont         = cmr5 at 10pt
\font \subhfont               = cmr10 scaled\magstep4
\font \tafonts                = cmbx7  scaled\magstep2
\font \tafontss               = cmbx5  scaled\magstep2
\font \tafontt                = cmbx10 scaled\magstep2
\font \tams                   = cmmib10
\font \tamss                  = cmmib10 scaled 700
\font \tamt                   = cmmib10 scaled\magstep2
\font \tass                   = cmsy7  scaled\magstep2
\font \tasss                  = cmsy5  scaled\magstep2
\font \tast                   = cmsy10 scaled\magstep2
\font \tasys                  = cmex10 scaled\magstep1
\font \tasyt                  = cmex10 scaled\magstep2
\font \tbfonts                = cmbx7  scaled\magstep1
\font \tbfontss               = cmbx5  scaled\magstep1
\font \tbfontt                = cmbx10 scaled\magstep1
\font \tbms                   = cmmib10 scaled 833
\font \tbmss                  = cmmib10 scaled 600
\font \tbmt                   = cmmib10 scaled\magstep1
\font \tbss                   = cmsy7  scaled\magstep1
\font \tbsss                  = cmsy5  scaled\magstep1
\font \tbst                   = cmsy10 scaled\magstep1
\font \tenbfne                = cmb10
\font \tensans                = cmss10
\font \tpfonts                = cmbx7  scaled\magstep3
\font \tpfontss               = cmbx5  scaled\magstep3
\font \tpfontt                = cmbx10 scaled\magstep3
\font \tpmt                   = cmmib10 scaled\magstep3
\font \tpss                   = cmsy7  scaled\magstep3
\font \tpsss                  = cmsy5  scaled\magstep3
\font \tpst                   = cmsy10 scaled\magstep3
\font \tpsyt                  = cmex10 scaled\magstep3
\vsize=19.3cm
\hsize=12.2cm
\hfuzz=2pt
\tolerance=500
\abovedisplayskip=3 mm plus6pt minus 4pt
\belowdisplayskip=3 mm plus6pt minus 4pt
\abovedisplayshortskip=0mm plus6pt minus 2pt
\belowdisplayshortskip=2 mm plus4pt minus 4pt
\predisplaypenalty=0
\clubpenalty=10000
\widowpenalty=10000
\frenchspacing
\newdimen\oldparindent\oldparindent=1.5em
\parindent=1.5em
\skewchar\ninei='177 \skewchar\sixi='177
\skewchar\ninesy='60 \skewchar\sixsy='60
\hyphenchar\ninett=-1
\def\newline{\hfil\break}%
\catcode`@=11
\def\folio{\ifnum\pageno<\z@
\uppercase\expandafter{\romannumeral-\pageno}%
\else\number\pageno \fi}
\catcode`@=12 
  \mathchardef\Gamma="0100
  \mathchardef\Delta="0101
  \mathchardef\Theta="0102
  \mathchardef\Lambda="0103
  \mathchardef\Xi="0104
  \mathchardef\Pi="0105
  \mathchardef\Sigma="0106
  \mathchardef\Upsilon="0107
  \mathchardef\Phi="0108
  \mathchardef\Psi="0109
  \mathchardef\Omega="010A
  \mathchardef\bfGamma="0\the\bffam 00
  \mathchardef\bfDelta="0\the\bffam 01
  \mathchardef\bfTheta="0\the\bffam 02
  \mathchardef\bfLambda="0\the\bffam 03
  \mathchardef\bfXi="0\the\bffam 04
  \mathchardef\bfPi="0\the\bffam 05
  \mathchardef\bfSigma="0\the\bffam 06
  \mathchardef\bfUpsilon="0\the\bffam 07
  \mathchardef\bfPhi="0\the\bffam 08
  \mathchardef\bfPsi="0\the\bffam 09
  \mathchardef\bfOmega="0\the\bffam 0A
\def\sun{\hbox{$\odot$}}
\def\la{\mathrel{\mathchoice {\vcenter{\offinterlineskip\halign{\hfil
$\displaystyle##$\hfil\cr<\cr\sim\cr}}}
{\vcenter{\offinterlineskip\halign{\hfil$\textstyle##$\hfil\cr<\cr\sim\cr}}}
{\vcenter{\offinterlineskip\halign{\hfil$\scriptstyle##$\hfil\cr<\cr\sim\cr}}}
{\vcenter{\offinterlineskip\halign{\hfil$\scriptscriptstyle##$\hfil\cr<\cr
\sim\cr}}}}}

\def\sq{\hbox{\rlap{$\sqcap$}$\sqcup$}}

\def\utw{\smash{\rlap{\lower5pt\hbox{$\sim$}}}}
\def\udtw{\smash{\rlap{\lower6pt\hbox{$\approx$}}}}

\def\diameter{{\ifmmode\mathchoice
{\ooalign{\hfil\hbox{$\displaystyle/$}\hfil\crcr
{\hbox{$\displaystyle\mathchar"20D$}}}}
{\ooalign{\hfil\hbox{$\textstyle/$}\hfil\crcr
{\hbox{$\textstyle\mathchar"20D$}}}}
{\ooalign{\hfil\hbox{$\scriptstyle/$}\hfil\crcr
{\hbox{$\scriptstyle\mathchar"20D$}}}}
{\ooalign{\hfil\hbox{$\scriptscriptstyle/$}\hfil\crcr
{\hbox{$\scriptscriptstyle\mathchar"20D$}}}}
\else{\ooalign{\hfil/\hfil\crcr\mathhexbox20D}}%
\fi}}


\def\bbbc{{\mathchoice {\setbox0=\hbox{$\displaystyle\rm C$}\hbox{\hbox
to0pt{\kern0.4\wd0\vrule height0.9\ht0\hss}\box0}}
{\setbox0=\hbox{$\textstyle\rm C$}\hbox{\hbox
to0pt{\kern0.4\wd0\vrule height0.9\ht0\hss}\box0}}
{\setbox0=\hbox{$\scriptstyle\rm C$}\hbox{\hbox
to0pt{\kern0.4\wd0\vrule height0.9\ht0\hss}\box0}}
{\setbox0=\hbox{$\scriptscriptstyle\rm C$}\hbox{\hbox
to0pt{\kern0.4\wd0\vrule height0.9\ht0\hss}\box0}}}}
\def\bbbe{{\mathchoice {\setbox0=\hbox{\smalletextfont e}\hbox{\raise
0.1\ht0\hbox to0pt{\kern0.4\wd0\vrule width0.3pt height0.7\ht0\hss}\box0}}
{\setbox0=\hbox{\smalletextfont e}\hbox{\raise
0.1\ht0\hbox to0pt{\kern0.4\wd0\vrule width0.3pt height0.7\ht0\hss}\box0}}
{\setbox0=\hbox{\smallescriptfont e}\hbox{\raise
0.1\ht0\hbox to0pt{\kern0.5\wd0\vrule width0.2pt height0.7\ht0\hss}\box0}}
{\setbox0=\hbox{\smallescriptscriptfont e}\hbox{\raise
0.1\ht0\hbox to0pt{\kern0.4\wd0\vrule width0.2pt height0.7\ht0\hss}\box0}}}}
\def\bbbq{{\mathchoice {\setbox0=\hbox{$\displaystyle\rm Q$}\hbox{\raise
0.15\ht0\hbox to0pt{\kern0.4\wd0\vrule height0.8\ht0\hss}\box0}}
{\setbox0=\hbox{$\textstyle\rm Q$}\hbox{\raise
0.15\ht0\hbox to0pt{\kern0.4\wd0\vrule height0.8\ht0\hss}\box0}}
{\setbox0=\hbox{$\scriptstyle\rm Q$}\hbox{\raise
0.15\ht0\hbox to0pt{\kern0.4\wd0\vrule height0.7\ht0\hss}\box0}}
{\setbox0=\hbox{$\scriptscriptstyle\rm Q$}\hbox{\raise
0.15\ht0\hbox to0pt{\kern0.4\wd0\vrule height0.7\ht0\hss}\box0}}}}
\def\bbbt{{\mathchoice {\setbox0=\hbox{$\displaystyle\rm
T$}\hbox{\hbox to0pt{\kern0.3\wd0\vrule height0.9\ht0\hss}\box0}}
{\setbox0=\hbox{$\textstyle\rm T$}\hbox{\hbox
to0pt{\kern0.3\wd0\vrule height0.9\ht0\hss}\box0}}
{\setbox0=\hbox{$\scriptstyle\rm T$}\hbox{\hbox
to0pt{\kern0.3\wd0\vrule height0.9\ht0\hss}\box0}}
{\setbox0=\hbox{$\scriptscriptstyle\rm T$}\hbox{\hbox
to0pt{\kern0.3\wd0\vrule height0.9\ht0\hss}\box0}}}}
\def\bbbs{{\mathchoice
{\setbox0=\hbox{$\displaystyle     \rm S$}\hbox{\raise0.5\ht0\hbox
to0pt{\kern0.35\wd0\vrule height0.45\ht0\hss}\hbox
to0pt{\kern0.55\wd0\vrule height0.5\ht0\hss}\box0}}
{\setbox0=\hbox{$\textstyle        \rm S$}\hbox{\raise0.5\ht0\hbox
to0pt{\kern0.35\wd0\vrule height0.45\ht0\hss}\hbox
to0pt{\kern0.55\wd0\vrule height0.5\ht0\hss}\box0}}
{\setbox0=\hbox{$\scriptstyle      \rm S$}\hbox{\raise0.5\ht0\hbox
to0pt{\kern0.35\wd0\vrule height0.45\ht0\hss}\raise0.05\ht0\hbox
to0pt{\kern0.5\wd0\vrule height0.45\ht0\hss}\box0}}
{\setbox0=\hbox{$\scriptscriptstyle\rm S$}\hbox{\raise0.5\ht0\hbox
to0pt{\kern0.4\wd0\vrule height0.45\ht0\hss}\raise0.05\ht0\hbox
to0pt{\kern0.55\wd0\vrule height0.45\ht0\hss}\box0}}}}
\def\bbbz{{\mathchoice {\hbox{$\sans\textstyle Z\kern-0.4em Z$}}
{\hbox{$\sans\textstyle Z\kern-0.4em Z$}}
{\hbox{$\sans\scriptstyle Z\kern-0.3em Z$}}
{\hbox{$\sans\scriptscriptstyle Z\kern-0.2em Z$}}}}
\def\qed{\ifmmode\sq\else{\unskip\nobreak\hfil
\penalty50\hskip1em\null\nobreak\hfil\sq
\parfillskip=0pt\finalhyphendemerits=0\endgraf}\fi}
\newfam\sansfam
\textfont\sansfam=\tensans\scriptfont\sansfam=\sevensans
\scriptscriptfont\sansfam=\fivesans
\def\sans{\fam\sansfam\tensans}
\def\stackfigbox{\if
Y\FIG\global\setbox\figbox=\vbox{\unvbox\figbox\box1}%
\else\global\setbox\figbox=\vbox{\box1}\global\let\FIG=Y\fi}
\def\placefigure{\dimen0=\ht1\advance\dimen0by\dp1
\advance\dimen0by5\baselineskip
\advance\dimen0by0.33333 cm
\ifdim\dimen0>\vsize\pageinsert\box1\vfill\endinsert
\else
\if Y\FIG\stackfigbox\else
\dimen0=\pagetotal\ifdim\dimen0<\pagegoal
\advance\dimen0by\ht1\advance\dimen0by\dp1\advance\dimen0by1.16666cm
\ifdim\dimen0>\pagegoal\stackfigbox
\else\box1\vskip3.33333 mm\fi
\else\box1\vskip3.33333 mm\fi\fi\fi}
%
\def\begfig#1cm#2\endfig{\par
\setbox1=\vbox{\dimen0=#1true cm\advance\dimen0
by0.83333 cm\kern\dimen0#2}\placefigure}
\def\begdoublefig#1cm #2 #3 \enddoublefig{\begfig#1cm%
\vskip-.8333\baselineskip\line{\vtop{\hsize=0.46\hsize#2}\hfill
\vtop{\hsize=0.46\hsize#3}}\endfig}
\def\begfigsidebottom#1cm#2cm#3\endfigsidebottom{\dimen0=#2true cm
\ifdim\dimen0<0.4\hsize\message{Room for legend to narrow;
begfigsidebottom changed to begfig}\begfig#1cm#3\endfig\else
\par\def\figure##1##2{\vbox{\noindent\petit{\bf
Fig.\ts##1\unskip.\ }\ignorespaces ##2\par}}%
\dimen0=\hsize\advance\dimen0 by-.66666 cm\advance\dimen0 by-#2true cm
\setbox1=\vbox{\hbox{\hbox to\dimen0{\vrule height#1true cm\hrulefill}%
\kern.66666 cm\vbox{\hsize=#2true cm#3}}}\placefigure\fi}
\def\begfigsidetop#1cm#2cm#3\endfigsidetop{\dimen0=#2true cm
\ifdim\dimen0<0.4\hsize\message{Room for legend to narrow; begfigsidetop
changed to begfig}\begfig#1cm#3\endfig\else
\par\def\figure##1##2{\vbox{\noindent\petit{\bf
Fig.\ts##1\unskip.\ }\ignorespaces ##2\par}}%
\dimen0=\hsize\advance\dimen0 by-.66666 cm\advance\dimen0 by-#2true cm
\setbox1=\vbox{\hbox{\hbox to\dimen0{\vrule height#1true cm\hrulefill}%
\kern.66666 cm\vbox to#1true cm{\hsize=#2true cm#3\vfill
}}}\placefigure\fi}
\def\figure#1#2{\vskip0.83333 cm\setbox0=\vbox{\noindent\petit{\bf
Fig.\ts#1\unskip.\ }\ignorespaces #2\smallskip
\count255=0\global\advance\count255by\prevgraf}%
\ifnum\count255>1\box0\else
\centerline{\petit{\bf Fig.\ts#1\unskip.\
}\ignorespaces#2}\smallskip\fi}
\def\tabcap#1#2{\smallskip\vbox{\noindent\petit{\bf Table\ts#1\unskip.\
}\ignorespaces #2\medskip}}
\def\begtab#1cm#2\endtab{\par
   \ifvoid\topins\midinsert\medskip\vbox{#2\kern#1true cm}\endinsert
   \else\topinsert\vbox{#2\kern#1true cm}\endinsert\fi}
\def\begpet{\vskip6pt\bgroup\petit}
\def\endpet{\vskip6pt\egroup}
\newcount\frpages
\newcount\frpagegoal
\def\freepage#1{\global\frpagegoal=#1\relax\global\frpages=0\relax
\loop\global\advance\frpages by 1\relax
\message{Doing freepage \the\frpages\space of
\the\frpagegoal}\null\vfill\eject
\ifnum\frpagegoal>\frpages\repeat}
\newdimen\refindent
\def\begrefchapter#1{\titlea{}{\ignorespaces#1}%
\bgroup\petit
\setbox0=\hbox{1000.\enspace}\refindent=\wd0}
\def\ref{\goodbreak
\hangindent\oldparindent\hangafter=1
\noindent\ignorespaces}
\def\refno#1{\goodbreak
\hangindent\refindent\hangafter=1
\noindent\hbox to\refindent{#1\hss}\ignorespaces}
\def\endref{\goodbreak\endpet}
\def\vec#1{{\textfont1=\tams\scriptfont1=\tamss
\textfont0=\tenbf\scriptfont0=\sevenbf
\mathchoice{\hbox{$\displaystyle#1$}}{\hbox{$\textstyle#1$}}
{\hbox{$\scriptstyle#1$}}{\hbox{$\scriptscriptstyle#1$}}}}
\def\petit{\def\rm{\fam0\ninerm}%
\textfont0=\ninerm \scriptfont0=\sixrm \scriptscriptfont0=\fiverm
 \textfont1=\ninei \scriptfont1=\sixi \scriptscriptfont1=\fivei
 \textfont2=\ninesy \scriptfont2=\sixsy \scriptscriptfont2=\fivesy
 \def\it{\fam\itfam\nineit}%
 \textfont\itfam=\nineit
 \def\sl{\fam\slfam\ninesl}%
 \textfont\slfam=\ninesl
 \def\bf{\fam\bffam\ninebf}%
 \textfont\bffam=\ninebf \scriptfont\bffam=\sixbf
 \scriptscriptfont\bffam=\fivebf
 \def\sans{\fam\sansfam\ninesans}%
 \textfont\sansfam=\ninesans \scriptfont\sansfam=\sixsans
 \scriptscriptfont\sansfam=\fivesans
 \def\tt{\fam\ttfam\ninett}%
 \textfont\ttfam=\ninett
 \normalbaselineskip=11pt
 \setbox\strutbox=\hbox{\vrule height7pt depth2pt width0pt}%
 \normalbaselines\rm
\def\vec##1{{\textfont1=\tbms\scriptfont1=\tbmss
\textfont0=\ninebf\scriptfont0=\sixbf
\mathchoice{\hbox{$\displaystyle##1$}}{\hbox{$\textstyle##1$}}
{\hbox{$\scriptstyle##1$}}{\hbox{$\scriptscriptstyle##1$}}}}}
\nopagenumbers
%
\let\header=Y
\let\FIG=N
\newbox\figbox
\output={\if N\header\headline={\hfil}\fi\plainoutput\global\let\header=Y
\if Y\FIG\topinsert\unvbox\figbox\endinsert\global\let\FIG=N\fi}
\let\lasttitle=N
\def\bookauthor#1{\vfill\eject
     \bgroup
     \baselineskip=22pt
     \lineskip=0pt
     \pretolerance=10000
     \authfont
     \rightskip 0pt plus 6em
     \centerpar{#1}\vskip1.66666 cm\egroup}
\def\bookhead#1#2{\bgroup
     \baselineskip=36pt
     \lineskip=0pt
     \pretolerance=10000
     \headfont
     \rightskip 0pt plus 6em
     \centerpar{#1}\vskip0.83333 cm
     \baselineskip=22pt
     \subhfont\centerpar{#2}\vfill
     \parindent=0pt
     \baselineskip=16pt
     \leftskip=1.83333cm
     \markfont Springer-Verlag\newline
     Berlin Heidelberg New York\newline
     London Paris Tokyo Singapore\bigskip\bigskip
     [{\it This is page III of your manuscript and will be reset by
     Springer.}]
     \egroup\let\header=N\eject}
\def\centerpar#1{{\parfillskip=0pt
\rightskip=0pt plus 1fil
\leftskip=0pt plus 1fil
\advance\leftskip by\oldparindent
\advance\rightskip by\oldparindent
\def\newline{\break}%
\noindent\ignorespaces#1\par}}
\def\part#1#2{\vfill\supereject\let\header=N
\centerline{\subhfont#1}%
\vskip75pt
     \bgroup
\textfont0=\tpfontt \scriptfont0=\tpfonts \scriptscriptfont0=\tpfontss
\textfont1=\tpmt \scriptfont1=\tbmt \scriptscriptfont1=\tams
\textfont2=\tpst \scriptfont2=\tpss \scriptscriptfont2=\tpsss
\textfont3=\tpsyt \scriptfont3=\tasys \scriptscriptfont3=\tenex
     \baselineskip=20pt
     \lineskip=0pt
     \pretolerance=10000
     \tpfontt
     \centerpar{#2}
     \vfill\eject\egroup\ignorespaces}
\newtoks\AUTHOR
\newtoks\HEAD
\catcode`\@=\active
\def\author#1{\bgroup
\baselineskip=22pt
\lineskip=0pt
\pretolerance=10000
\markfont
\centerpar{#1}\bigskip\egroup
{\def@##1{}%
\setbox0=\hbox{\petit\kern2.08333 cc\ignorespaces#1\unskip}%
\ifdim\wd0>\hsize
\message{The names of the authors exceed the headline, please use a }%
\message{short form with AUTHORRUNNING}\gdef\leftheadline{%
\hbox to2.08333 cc{\folio\hfil}AUTHORS suppressed due to excessive
length\hfil}%
\global\AUTHOR={AUTHORS were to long}\else
\xdef\leftheadline{\hbox to2.08333
cc{\noexpand\folio\hfil}\ignorespaces#1\hfill}%
\global\AUTHOR={\def@##1{}\ignorespaces#1\unskip}\fi
}\let\INS=E}
\def\address#1{\bgroup
\centerpar{#1}\bigskip\egroup
\catcode`\@=12
\vskip2cm\noindent\ignorespaces}
\let\INS=N%
\def@#1{\if N\INS\unskip\ $^{#1}$\else\if
E\INS\noindent$^{#1}$\let\INS=Y\ignorespaces
\else\par
\noindent$^{#1}$\ignorespaces\fi\fi}%
\catcode`\@=12
\headline={\petit\def\newline{ }\def\fonote#1{}\ifodd\pageno
\rightheadline\else\leftheadline\fi}
\def\rightheadline{\hfil Missing CONTRIBUTION
title\hbox to2.08333 cc{\hfil\folio}}
\def\leftheadline{\hbox to2.08333 cc{\folio\hfil}Missing name(s) of the
author(s)\hfil}
\nopagenumbers
\let\header=Y

\let\lasttitle=N
 \def\contribution#1{\vfill\supereject
 \ifodd\pageno\else\null\vfill\supereject\fi
 \let\header=N\bgroup
 \textfont0=\tafontt \scriptfont0=\tafonts \scriptscriptfont0=\tafontss
 \textfont1=\tamt \scriptfont1=\tams \scriptscriptfont1=\tams
 \textfont2=\tast \scriptfont2=\tass \scriptscriptfont2=\tasss
 \par\baselineskip=16pt
     \lineskip=16pt
     \tafontt
     \raggedright
     \pretolerance=10000
     \noindent
     \centerpar{\ignorespaces#1}%
     \vskip12pt\egroup
     \nobreak
     \parindent=0pt
     \everypar={\global\parindent=1.5em
     \global\let\lasttitle=N\global\everypar={}}%
     \global\let\lasttitle=A%
     \setbox0=\hbox{\petit\def\newline{ }\def\fonote##1{}\kern2.08333
     cc\ignorespaces#1}\ifdim\wd0>\hsize
     \message{Your CONTRIBUTIONtitle exceeds the headline,
please use a short form
with CONTRIBUTIONRUNNING}\gdef\rightheadline{\hfil CONTRIBUTION title
suppressed due to excessive length\hbox to2.08333 cc{\hfil\folio}}%
\global\HEAD={HEAD was to long}\else
\gdef\rightheadline{\hfill\ignorespaces#1\unskip\hbox to2.08333
cc{\hfil\folio}}\global\HEAD={\ignorespaces#1\unskip}\fi
\catcode`\@=\active
     \ignorespaces}
 \def\contributionnext#1{\vfill\supereject
 \let\header=N\bgroup
 \textfont0=\tafontt \scriptfont0=\tafonts \scriptscriptfont0=\tafontss
 \textfont1=\tamt \scriptfont1=\tams \scriptscriptfont1=\tams
 \textfont2=\tast \scriptfont2=\tass \scriptscriptfont2=\tasss
 \par\baselineskip=16pt
     \lineskip=16pt
     \tafontt
     \raggedright
     \pretolerance=10000
     \noindent
     \centerpar{\ignorespaces#1}%
     \vskip12pt\egroup
     \nobreak
     \parindent=0pt
     \everypar={\global\parindent=1.5em
     \global\let\lasttitle=N\global\everypar={}}%
     \global\let\lasttitle=A%
     \setbox0=\hbox{\petit\def\newline{ }\def\fonote##1{}\kern2.08333
     cc\ignorespaces#1}\ifdim\wd0>\hsize
     \message{Your CONTRIBUTIONtitle exceeds the headline,
please use a short form
with CONTRIBUTIONRUNNING}\gdef\rightheadline{\hfil CONTRIBUTION title
suppressed due to excessive length\hbox to2.08333 cc{\hfil\folio}}%
\global\HEAD={HEAD was to long}\else
\gdef\rightheadline{\hfill\ignorespaces#1\unskip\hbox to2.08333
cc{\hfil\folio}}\global\HEAD={\ignorespaces#1\unskip}\fi
\catcode`\@=\active
     \ignorespaces}
\def\motto#1#2{\bgroup\petit\leftskip=5.41666cm\noindent\ignorespaces#1
\if!#2!\else\medskip\noindent\it\ignorespaces#2\fi\bigskip\egroup
\let\lasttitle=M
\parindent=0pt
\everypar={\global\parindent=\oldparindent
\global\let\lasttitle=N\global\everypar={}}%
\global\let\lasttitle=M%
\ignorespaces}
\def\abstract#1{\bgroup\petit\noindent
{\bf Abstract: }\ignorespaces#1\vskip28pt\egroup
\let\lasttitle=N
\parindent=0pt
\everypar={\global\parindent=\oldparindent
\global\let\lasttitle=N\global\everypar={}}%
\ignorespaces}
\def\titlea#1#2{\if N\lasttitle\else\vskip-28pt
     \fi
     \vskip18pt plus 4pt minus4pt
     \bgroup
\textfont0=\tbfontt \scriptfont0=\tbfonts \scriptscriptfont0=\tbfontss
\textfont1=\tbmt \scriptfont1=\tbms \scriptscriptfont1=\tbmss
\textfont2=\tbst \scriptfont2=\tbss \scriptscriptfont2=\tbsss
\textfont3=\tasys \scriptfont3=\tenex \scriptscriptfont3=\tenex
     \baselineskip=16pt
     \lineskip=0pt
     \pretolerance=10000
     \noindent
     \tbfontt
     \rightskip 0pt plus 6em
     \setbox0=\vbox{\vskip23pt\def\fonote##1{}%
     \noindent
     \if!#1!\ignorespaces#2
     \else\setbox0=\hbox{\ignorespaces#1\unskip\ }\hangindent=\wd0
     \hangafter=1\box0\ignorespaces#2\fi
     \vskip18pt}%
     \dimen0=\pagetotal\advance\dimen0 by-\pageshrink
     \ifdim\dimen0<\pagegoal
     \dimen0=\ht0\advance\dimen0 by\dp0\advance\dimen0 by
     3\normalbaselineskip
     \advance\dimen0 by\pagetotal
     \ifdim\dimen0>\pagegoal\eject\fi\fi
     \noindent
     \if!#1!\ignorespaces#2
     \else\setbox0=\hbox{\ignorespaces#1\unskip\ }\hangindent=\wd0
     \hangafter=1\box0\ignorespaces#2\fi
     \vskip18pt plus4pt minus4pt\egroup
     \nobreak
     \parindent=0pt
     \everypar={\global\parindent=\oldparindent
     \global\let\lasttitle=N\global\everypar={}}%
     \global\let\lasttitle=A%
     \ignorespaces}
 \def\titleb#1#2{\if N\lasttitle\else\vskip-28pt
     \fi
     \vskip18pt plus 4pt minus4pt
     \bgroup
\textfont0=\tenbf \scriptfont0=\sevenbf \scriptscriptfont0=\fivebf
\textfont1=\tams \scriptfont1=\tamss \scriptscriptfont1=\tbmss
     \lineskip=0pt
     \pretolerance=10000
     \noindent
     \bf
     \rightskip 0pt plus 6em
     \setbox0=\vbox{\vskip23pt\def\fonote##1{}%
     \noindent
     \if!#1!\ignorespaces#2
     \else\setbox0=\hbox{\ignorespaces#1\unskip\enspace}\hangindent=\wd0
     \hangafter=1\box0\ignorespaces#2\fi
     \vskip10pt}%
     \dimen0=\pagetotal\advance\dimen0 by-\pageshrink
     \ifdim\dimen0<\pagegoal
     \dimen0=\ht0\advance\dimen0 by\dp0\advance\dimen0 by
     3\normalbaselineskip
     \advance\dimen0 by\pagetotal
     \ifdim\dimen0>\pagegoal\eject\fi\fi
     \noindent
     \if!#1!\ignorespaces#2
     \else\setbox0=\hbox{\ignorespaces#1\unskip\enspace}\hangindent=\wd0
     \hangafter=1\box0\ignorespaces#2\fi
     \vskip8pt plus4pt minus4pt\egroup
     \nobreak
     \parindent=0pt
     \everypar={\global\parindent=\oldparindent
     \global\let\lasttitle=N\global\everypar={}}%
     \global\let\lasttitle=B%
     \ignorespaces}
 \def\titlec#1#2{\if N\lasttitle\else\vskip-23pt
     \fi
     \vskip18pt plus 4pt minus4pt
     \bgroup
\textfont0=\tenbfne \scriptfont0=\sevenbf \scriptscriptfont0=\fivebf
\textfont1=\tams \scriptfont1=\tamss \scriptscriptfont1=\tbmss
     \tenbfne
     \lineskip=0pt
     \pretolerance=10000
     \noindent
     \rightskip 0pt plus 6em
     \setbox0=\vbox{\vskip23pt\def\fonote##1{}%
     \noindent
     \if!#1!\ignorespaces#2
     \else\setbox0=\hbox{\ignorespaces#1\unskip\enspace}\hangindent=\wd0
     \hangafter=1\box0\ignorespaces#2\fi
     \vskip6pt}%
     \dimen0=\pagetotal\advance\dimen0 by-\pageshrink
     \ifdim\dimen0<\pagegoal
     \dimen0=\ht0\advance\dimen0 by\dp0\advance\dimen0 by
     2\normalbaselineskip
     \advance\dimen0 by\pagetotal
     \ifdim\dimen0>\pagegoal\eject\fi\fi
     \noindent
     \if!#1!\ignorespaces#2
     \else\setbox0=\hbox{\ignorespaces#1\unskip\enspace}\hangindent=\wd0
     \hangafter=1\box0\ignorespaces#2\fi
     \vskip6pt plus4pt minus4pt\egroup
     \nobreak
     \parindent=0pt
     \everypar={\global\parindent=\oldparindent
     \global\let\lasttitle=N\global\everypar={}}%
     \global\let\lasttitle=C%
     \ignorespaces}
 \def\titled#1{\if N\lasttitle\else\vskip-\baselineskip
     \fi
     \vskip12pt plus 4pt minus 4pt
     \bgroup
\textfont1=\tams \scriptfont1=\tamss \scriptscriptfont1=\tbmss
     \bf
     \noindent
     \ignorespaces#1\ \ignorespaces\egroup
     \ignorespaces}
\let\ts=\thinspace
\def\footnoterule{\kern-3pt\hrule width 1.66666 cm\kern2.6pt}
\newcount\footcount \footcount=0
\def\advftncnt{\advance\footcount by1\global\footcount=\footcount}
\def\fonote#1{\advftncnt$^{\the\footcount}$\begingroup\petit
\parfillskip=0pt plus 1fil
\def\textindent##1{\hangindent0.5\oldparindent\noindent\hbox
to0.5\oldparindent{##1\hss}\ignorespaces}%
\vfootnote{$^{\the\footcount}$}{#1\vskip-9.69pt}\endgroup}
\def\item#1{\par\noindent
\hangindent6.5 mm\hangafter=0
\llap{#1\enspace}\ignorespaces}

\def\titleao#1{\vfill\supereject
\ifodd\pageno\else\null\vfill\supereject\fi
\let\header=N
     \bgroup
\textfont0=\tafontt \scriptfont0=\tafonts \scriptscriptfont0=\tafontss
\textfont1=\tamt \scriptfont1=\tams \scriptscriptfont1=\tamss
\textfont2=\tast \scriptfont2=\tass \scriptscriptfont2=\tasss
\textfont3=\tasyt \scriptfont3=\tasys \scriptscriptfont3=\tenex
     \baselineskip=18pt
     \lineskip=0pt
     \pretolerance=10000
     \tafontt
     \centerpar{#1}%
     \vskip75pt\egroup
     \nobreak
     \parindent=0pt
     \everypar={\global\parindent=\oldparindent
     \global\let\lasttitle=N\global\everypar={}}%
     \global\let\lasttitle=A%
     \ignorespaces}






\def\leaderfill{\kern0.5em\leaders\hbox to 0.5em{\hss.\hss}\hfill\kern
0.5em}
\newdimen\chapindent
\newdimen\sectindent
\newdimen\subsecindent
\newdimen\thousand
\setbox0=\hbox{\bf 10. }\chapindent=\wd0
\setbox0=\hbox{10.10 }\sectindent=\wd0
\setbox0=\hbox{10.10.1 }\subsecindent=\wd0
\setbox0=\hbox{\enspace 100}\thousand=\wd0
\def\contpart#1#2{\medskip\noindent
\vbox{\kern10pt\leftline{\textfont1=\tams
\scriptfont1=\tamss\scriptscriptfont1=\tbmss\bf
\advance\chapindent by\sectindent
\hbox to\chapindent{\ignorespaces#1\hss}\ignorespaces#2}\kern8pt}%
\let\lasttitle=Y\par}
\def\contcontribution#1#2{\if N\lasttitle\bigskip\fi
\let\lasttitle=N\line{{\textfont1=\tams
\scriptfont1=\tamss\scriptscriptfont1=\tbmss\bf#1}%
\if!#2!\hfill\else\leaderfill\hbox to\thousand{\hss#2}\fi}\par}
\def\conttitlea#1#2#3{\line{\hbox to
\chapindent{\strut\bf#1\hss}{\textfont1=\tams
\scriptfont1=\tamss\scriptscriptfont1=\tbmss\bf#2}%
\if!#3!\hfill\else\leaderfill\hbox to\thousand{\hss#3}\fi}\par}
\def\conttitleb#1#2#3{\line{\kern\chapindent\hbox
to\sectindent{\strut#1\hss}{#2}%
\if!#3!\hfill\else\leaderfill\hbox to\thousand{\hss#3}\fi}\par}
\def\conttitlec#1#2#3{\line{\kern\chapindent\kern\sectindent
\hbox to\subsecindent{\strut#1\hss}{#2}%
\if!#3!\hfill\else\leaderfill\hbox to\thousand{\hss#3}\fi}\par}
\long\def\lemma#1#2{\removelastskip\vskip\baselineskip\noindent{\tenbfne
Lemma\if!#1!\else\ #1\fi\ \ }{\it\ignorespaces#2}\vskip\baselineskip}
\long\def\proposition#1#2{\removelastskip\vskip\baselineskip\noindent{\tenbfne
Proposition\if!#1!\else\ #1\fi\ \ }{\it\ignorespaces#2}\vskip\baselineskip}
\long\def\theorem#1#2{\removelastskip\vskip\baselineskip\noindent{\tenbfne
Theorem\if!#1!\else\ #1\fi\ \ }{\it\ignorespaces#2}\vskip\baselineskip}
\long\def\corollary#1#2{\removelastskip\vskip\baselineskip\noindent{\tenbfne
Corollary\if!#1!\else\ #1\fi\ \ }{\it\ignorespaces#2}\vskip\baselineskip}
\long\def\example#1#2{\removelastskip\vskip\baselineskip\noindent{\tenbfne
Example\if!#1!\else\ #1\fi\ \ }\ignorespaces#2\vskip\baselineskip}
\long\def\exercise#1#2{\removelastskip\vskip\baselineskip\noindent{\tenbfne
Exercise\if!#1!\else\ #1\fi\ \ }\ignorespaces#2\vskip\baselineskip}
\long\def\problem#1#2{\removelastskip\vskip\baselineskip\noindent{\tenbfne
Problem\if!#1!\else\ #1\fi\ \ }\ignorespaces#2\vskip\baselineskip}
\long\def\solution#1#2{\removelastskip\vskip\baselineskip\noindent{\tenbfne
Solution\if!#1!\else\ #1\fi\ \ }\ignorespaces#2\vskip\baselineskip}


\long\def\definition#1#2{\removelastskip\vskip\baselineskip\noindent{\tenbfne
Definition\if!#1!\else\
#1\fi\ \ }\ignorespaces#2\vskip\baselineskip}
\def\frame#1{\bigskip\vbox{\hrule\hbox{\vrule\kern5pt
\vbox{\kern5pt\advance\hsize by-10.8pt
\centerline{\vbox{#1}}\kern5pt}\kern5pt\vrule}\hrule}\bigskip}
\def\frameddisplay#1#2{$$\vcenter{\hrule\hbox{\vrule\kern5pt
\vbox{\kern5pt\hbox{$\displaystyle#1$}%
\kern5pt}\kern5pt\vrule}\hrule}\eqno#2$$}
\def\typeset{\petit\noindent This book was processed by the author using
the \TeX\ macro package from Springer-Verlag.\par}
\outer\def\byebye{\bigskip\bigskip\typeset
\footcount=1\ifx\speciali\undefined\else
\loop\smallskip\noindent special character No\number\footcount:
\csname special\romannumeral\footcount\endcsname
\advance\footcount by 1\global\footcount=\footcount
\ifnum\footcount<11\repeat\fi
\gdef\leftheadline{\hbox to2.08333 cc{\folio\hfil}\ignorespaces
\the\AUTHOR\unskip: \the\HEAD\hfill}\vfill\supereject\end}
\contribution{Lithium and the Nature of  Brown Dwarf Candidates}
\author{Rafael Rebolo@1, Antonio Magazz\`u@2, Eduardo L. Mart\'\i n@{3}}
\address{@1 Instituto de Astrofisica de Canarias,
Via Lactea s/n, 38200 La Laguna, Tenerife, Spain
@2 Osservatorio Astrofisico di Catania, Citt\`a Universitaria, I-95125
Catania,Italy
@3 Astronomical Institute ``Anton Pannekoek'', Kruislaan 403, 1098 SJ,
Amsterdam, The Netherlands}

\abstract{This paper discusses the ability of Li to confirm the substellar
nature of a brown dwarf candidate.  Theoretical computations  using different
interior models agree that brown dwarfs with  masses below $\sim$ 0.065
M$_{\odot}$ preserve a significant fraction of their initial Li content while
for higher masses  total Li depletion occurs in very short timescales. Refined
spectral synthesis using  brown dwarf model atmospheres  shows that the Li I
resonance line at 670.8 nm should produce a conspicuous feature at effective
temperatures higher than 2000 K, suitable for spectroscopic detection in
present brown dwarf candidates.  We summarize the results of our  search for Li
in many of  the best candidates known so far.}

\titlea{1} {Introduction}

The $^7$Li isotope, the most abundant of the two stable lithium isotopes in
Nature, has been detected in hundreds of stars and in the Interstellar Medium.
The nuclear binding energy of $^7$Li nuclei is rather low, as a consequence,
these are efficiently destroyed in stellar interiors via proton collisions
$^7$Li(p,$\alpha$)$^4$He at relatively low temperatures, $\sim 2 \times 10^6$K.
The behaviour of the cross-section for this reaction against the temperature of
the plasma  allows to define a $^7$Li burning temperature such that at higher
temperatures the stellar plasma is essentially Li free. This burning
temperature has a weak dependence on the density conditions of the interiors,
varying in the range 1.8-2.4 $\times 10^6$ K for objects close to  the
substellar limit.  In this paper we deal with the ability of $^7$Li nuclei,
hereafter Li, to allow distinguishing between very low mass stars and massive
brown dwarfs (BD), not only on theoretical grounds,  but also observationally
through detection of Li features in the spectrum. Many of the ideas and results
presented here can be found in Rebolo, Mart\'\i n and Magazz\`u (1992),
Magazz\`u, Mart\'\i n and Rebolo (1993) and Mart\'\i n, Rebolo and Magazz\`u
(1994).

The present abundance of Li in our Galaxy is Li/H=10$^{-9}$, usually expressed
as $\log N({\rm Li}) = 3$ in the scale $\log N({\rm H}) = 12$. Such present
``cosmic'' abundance is derived from observations of T~Tauri stars, F stars in
young stellar clusters like Pleiades and Hyades, and Interstellar Medium. On
the other hand, observations of halo dwarfs show that the Li content in the
material from which our Galaxy formed was at least $\log N({\rm Li}) = 2$
(Spite \& Spite 1982, Rebolo et al. 1988). In fact, Li is one of the few
nuclear species produced in the first minutes of the life of the Universe
(Wagoner et al.\ 1967, Boesgaard \& Steigman 1985). Although there is still
controversy  on the exact abundance that came out from the Big Bang, it is
possible to claim that any star or substellar object formed in our Galaxy was
born with a lithium abundance higher than $\log N({\rm Li}) = 2$. This is
important in order to establish the existence of brown dwarfs.

\titlea{2}{Lithium Destruction at the Bottom of the Main Sequence}

The fragility of Li nuclei in stellar interiors makes possible to use this
element as a tracer of internal structure in stars of different type (see e.g.\
reviews by Michaud \& Charbonneau 1991, Rebolo 1991). Since about 30 years it
is known that Li is destroyed very efficiently in the interior of solar type
stars, but only recently it has been possible to estimate how efficient  this
destruction is in very late type stars.   Observations of late K and M pre-main
sequence (PMS) stars (Magazz\`u, Rebolo and Pavlenko 1992, Mart\'\i n et al.
1994a)   show that significant Li destruction  takes place during the early
evolution of PMS low mass  stars.  Early M-type  stars in the Pleiades (age 70
Myr) have reduced their initial Li content more than a factor 1000 (Garc\'\i a
L\'opez, Rebolo and Mart\'\i n 1994), even younger M3-M6 stars in the
$\alpha$~Per open cluster (age 50~Myr) have destroyed Li by similar factors
(Zapatero-Osorio et al., this volume).  The absence of Li  in $\alpha$ Per
M5-M6 stars -- masses in the range 0.2-0.1 $M\sun$ -- is the strongest
observational indication of the high  efficiency of very low mass stars in
destroying their initial Li content and  is  in good agreement with the
theoretical  predictions of D'Antona \& Mazzitelli (1994).  As far as we know,
the only very late M star retaining lithium in its atmosphere is UX~Tau~C
(Magazz\`u et al.\ 1991), the faintest object of the UX~Tau PMS system with
mass lower than 0.2 $M\sun$.  This star presents a strong Li absorption
resonance doublet, indicating that has suffered little destruction. However,
the age uf the UX~Tau system is only about 2 Myr! It can be inferred that very
low mass stars  with age greater than a few Myr have destroyed  a large
fraction of their initial Li.

\titlea{3}{The Lithium Test}

Visual inspection of the curves showing the evolution of central temperatures
versus time for very low mass stars and brown dwarfs (see e.g.\ D'Antona and
Mazzitelli 1985, Burrows and Liebert 1993) clearly denotes that  objects with
masses well below the substellar limit cannot reach the Li burning temperature.
Opposite to very low mass stars, these brown dwarfs must preserve a significant
amount of their initial Li content. The effective atmospheric temperatures of
objects close to the substellar limit are subject of controversy and difficult
to predict on theoretical grounds, however the available spectral type-$T_{\rm
eff}$ calibrations give effective temperatures of about 2500 K for the present
brown dwarf candidates, not very different to those of the coolest T Tauri
stars with detected Li. These considerations prompted Rebolo et al.\ (1992) to
compute the formation of the Li I resonance doublet at 670.8 nm  using Allard
(1990) brown dwarf model atmospheres with $T_{\rm eff}$ 2000 to
2700 K. The computations showed the formation of a very strong line (equivalent
width of several \AA) in such cool atmospheres and moved us
to suggest  its detection as a powerful tool to confirm the substellar
nature of brown dwarf candidates: the Li test (Rebolo et al. 1992)

We summarize in the following some recent results that support the
applicability of the test.  Both, regarding the
destruction/preservation of Li around the substellar limit  and the
formation of Li lines in the atmospheres of very cool objects.

  \titleb{3.1} {The Mass Limit for Preservation of Lithium}

A simple approach considering the structure of a brown dwarf as a $n = 3/2$
polytrope indicates that the central temperature of such an object never
reaches the value necessary to burn lithium for $M \la 0.07~M_\odot$.
Calculations of Li depletion in substellar objects using detailed interior
models show that Li  is preserved below $\sim$ 0.06~$M\sun$ (Magazz\`u et al.
1993).  In fact, Pozio (1991), Stringfellow (1989) (quoted in Bessell and
Stringfellow 1993)  and Magazz\'u et al. had  independently  considered this
problem and obtained  similar results.  More recently, Nelson et al. (1993) and
D'Antona \& Mazzitelli (1994) find essentially the same, Li is preserved below
$\sim$ 0.065~$M\sun$.  It is remarkable the good agreement between all these
works despite the use of different interior models, opacities and screening
factors. The mass limit for Li preservation is clearly below the substellar
mass limit, usually accepted to lay  between 0.08 and 0.07~$M\sun$.

In Fig. 1 we show the Li destruction curves from Magazz\`u et al. (1993). They
illustrate, similarly to those obtained by other authors, an interesting
additional point. For objects with masses 0.08-0.07 $M\sun$ it is predicted
total Li destruction, but this takes place only after $\sim$ 100 Myr at 0.08
M\sun~ and $\sim$ 200 Myr at 0.07 M\sun.  At the age of the Pleiades, these
objects should also show Li in their atmospheres, the mass limit for Li
preservation  coincides here with the hydrogen burning mass limit. The shape of
the destruction curves allow to predict a sharp transition between Li-poor and
Li-rich objects at the bottom of the MS of a cluster like the Pleiades.
Observations of fainter and  fainter objects  in young open clusters should
lead to a sudden  detection of Li at a given luminosity and spectral type,
which in turn would define an empirical location of the substellar mass limit.
In an older cluster like the Hyades (600 Myr) the luminosity of objects at the
substellar limit would be slightly higher than that of objects preserving  Li.

\begfig 8 cm
\figure{1}{Depletion of Li for masses close to the substellar limit, after
Magazz\`u et al. 1993}
\endfig

We have seen above that all theoretical considerations indicate that
lithium should be preserved below $\sim 0.065~M\sun$. There is
indication that lithium -- if present -- can be detected
spectroscopically at the physical conditions of the atmospheres of such
objects.

\titleb{3.2} {Detectability of Lithium in Brown Dwarfs.}

The formation of atomic and molecular Li lines in the spectrum of brown
dwarfs is a problem requiring careful modelling. So far, we have
investigated optical and infrared Li I lines using brown dwarf model
atmospheres with $T_{\rm eff}$ higher than 1800 K. In a near future we
plan to investigate the formation of lines from molecular species
containing Li atoms. The computations by Rebolo et al. (1992) and
 Magazz\`u et al.
(1993) clearly revealed the formation of strong Li lines at 670.8 nm in
high gravity (log g=5.0) low temperature (2000 K) models showing that
at these temperatures the Li absorption is very strong and well
detectable in the forest of molecular bands  present in the spectrum.
It was argued following Tsuji (1973) that the formation of molecules
containing Li would be relevant at temperatures lower than 1500 K and
that no significant effect on the formation of the resonance doublet
were expected at higher T$_{\rm eff}$. Recently,  Pavlenko et
al.\ (1994) analyzed in detail the LTE and NLTE formation of Li I lines
using Allard (1990) model atmospheres of T$_{\rm eff}$ 2000, 2500 and
3000 K.  In these new computations, the dissociation equilibria for
seven Li molecules (LiH, LiO, LiCl, LiF, LiBr, LiI, LiOH) was taken
into account, and also considered  atomic-molecular line lists
available in the Li resonance line region. More than 20 atoms in two
ionization states and 54 molecules  were considered in the state
equation system. The synthetic spectra were able to reproduce the TiO bands
in the region around
the Li resonance line, correctly describing positions and intensities
of the observed bands. The resulting LTE computations show the
formation of prominent Li I 670.8 nm lines (see Fig. 2), with
equivalent widths of several \AA. The NLTE effects are found to be
small, less than 0.1 dex in abundance. LTE and NLTE  curves of growth for the
weaker  Li I lines at 610.3 and 812.6 nm. are also given in Pavlenko et al.\
(1994).

\begfig 8 cm
\figure{2}{Spectral synthesis of the Li I resonance line (from Pavlenko et
al. 1994) obtained  using a brown dwarf model atmosphere of $T_{\rm eff}$=2500
and log g=5. Line profiles for log N(Li)=3.0, 2.0, 1.0, 0.0, -1.0 and -2.0 are
shown}
\endfig

Computations of  the most promising IR Li I lines also confirm the
resonance doublet as the most suitable feature for detecting  atomic Li
in brown dwarfs. Such detection is feasible for faint objects (R$\sim$22)
with the new generation of  large 8-10~m telescopes using intermediate
resolution spectroscopy (3-4 \AA).

\titlea{4} {Searching for Li in Brown Dwarf Candidates}

The first Li search in BD candidates was performed by Magazz\`u et al. (1993).
These authors observed a few relatively bright objects (GL 234 B, GL
473AB, GL 569B, etc) with high resolution spectroscopy (FWHM=0.4-0.8
\AA ). Since, as explained above, the Li lines are expected to be very
strong, Mart\'\i n et al. (1994b) observed fainter objects (down to
V=20.5) with intermediate resolution (FWHM=2-4 \AA) using 4m class
telescopes at La Palma and La Silla observatories. The Keck 10m
telescope has allowed to reach similar faint magnitudes, but with a
much higher resolution (FWHM=0.2 \AA , Marcy et al. 1994). None of
these three works has been able to report a Li detection.

Brown dwarf candidates can be found in three different contexts; in
binaries, in clusters, or free floating in the field. In the following,
we discuss the Li non-detections in each of these contexts.

\titleb{4.1} {Very Low Mass Binaries}

The best way of determining a substellar mass is to obtain a complete
orbital solution of a binary system containing a brown dwarf.  However,
this has not yet been possible, and presently we only have approximate
solutions for a handful of systems. Given the uncertainties some known
secondaries could have masses below the substellar limit.  In Table~1
we summarize the Li searches in BD candidate secondaries.  The upper
limits to the LiI equivalent widths (EW) come from Magazz\`u et al.
(1993) and Mart\'\i n et al. (1994b), and have been corrected for the
flux contribution of the primary.  The lower limits to the masses
derived from the Li non-detection come from comparison with the
predictions by Nelson et al.  (1993). The dynamical masses come from
the compilation made by Burrows et al. (1989), and Ianna et al. (1988)
for LHS 1047B. However, we note that the dynamical masses of GL 473AB
are probably much larger (see Henry, this volume), and the error bar
for LHS 1047B should be increased according to Henry \& McCarthy
(1993).
{}From the comparison of Li masses and dynamical masses in Table 1 we
find that there is consistency within the error bars. None of these
secondaries is a BD with mass less than about $0.065 M\sun$, but it
is not ruled out that some could have masses very close to the
substellar mass limit  (0.08 $M\sun$). Presently, neither the Li
test, nor the precision of orbital parameters can discriminate between
secondaries slightly above or below the substellar limit.
\vskip 1 truecm
\tabcap{1}{Li searches in VLM binaries}
\settabs\+& Nameaaaaaaaaaa&aaaaaaaa&aaaaaaaaaaaaaa&aaaaaaaaaaa&aaaaaaaaaaa\cr
\+&\cr
\hrule
\hrule
\+&\cr
\+& Object & Sp.T. & LiI (m\AA) & M/M$_\odot$ & M/M$_\odot$ \cr
\+&  &  &  & Lithium & Dynamical \cr
\+&\cr
\hrule
\+&\cr
\+&GL 65A,B  & M6 & $<$100  &  $>$0.065 & (0.115, 0.109)$\pm$0.008  \cr
\+&GL 234B  & M6: & $<$500  &  $>$0.065 & 0.08$\pm$0.01 \cr
\+&GL 473AB  & M6: & $<$500  &  $>$0.065 & (0.059, 0.051)$\pm$0.01 \cr
\+&GL 623B  & M6: & $<$700  &  $>$0.065 & 0.114$\pm$0.042 \cr
\+&LHS 1047B  & M6.5:  & $<$500  &  $>$0.066 & 0.055$\pm$0.032 \cr
\+&\cr
\hrule
\hrule
\smallskip
\noindent
Note: the dynamical masses for GL 473AB and LHS 1047B are controversial
(see text).
\vskip 0.1 truecm

\titleb{4.2}  {Young Open Clusters}

Several surveys for BD in nearby open clusters have been carried out in
the last few years (e.g.\ Jameson \& Skillen 1989, Simons \& Becklin
1992, Bryja et al.\  1994). Dozens of BD candidates were identified on
the basis of very red optical or IR colours, but for only very few of
them do proper motion and spectroscopic studies exist.  The Li test has
been applied to very few of these candidates:  the one discovered by
Rebolo et al. (1992) in the $\alpha$Per cluster, three of the PM
candidates in the Pleiades (Hambly et al.\ 1993), and the faintest
proper motion Hyades member in the sample of Bryja et al. (1994).

The Li test provides stronger mass constraints in $\alpha$Per and the
Pleiades  than in the Hyades because of the difference in ages of about
a factor ten.  The substellar Li region (SLR), the region of the HR diagram
where substellar objects preserving Li lay,  has an upper mass limit of
0.08~$M_\odot$ or
higher for ages lower than about 10$^8$ yr, while for ages of a few
$10^8$ yr the upper mass limit is $\sim 0.065~M_\odot$ (Mart\'\i n et
al.\ 1994b). Using the conversions from I-K colour to BC and $T_{\rm
eff}$ of Bessell (1991), all the BD candidates of Table 2 in
$\alpha$Per and the Pleiades fall inside the SLR, which is inconsistent
with the large Li depletion observed.  In order to bring the positions
of these stars outside the SLR, they would have to be shifted by 200~K
to hotter T$_{\rm eff}$. Current uncertainties in assigning
temperatures to these objects could account for this shift.

The Li mass constraints (summarized in Table 2) imply that AP 0323, HHJ
3, 10 and 14 are not BDs. Fainter and redder objects than these have
been detected towards the Pleiades (see Jameson in this volume), but
their membership to the cluster needs to be confirmed before the
decisive Li test is attempted.
\smallskip
\smallskip
\smallskip
\vskip 1 truecm
\tabcap{2}{Li searches in open cluster BD candidates}
\settabs\+&
Nameaaaaaaaaa&aaaaaaaa&aaaaaaaaaaaaaa&aaaaaaaaaaa&aaaaaaaaaaa&aaaaaaaaaa\cr
\+&\cr
\hrule
\hrule
\+&\cr
\+& Object & Sp.T. & LiI (m\AA) & M/M$_\odot$ & Cluster & Ref. \cr
\+&\cr
\hrule
\+&\cr
\+&AP 0323+48 &  M6 & $<$440 & $>$0.08  & $\alpha$ Per & ZRMG \cr
\+&HHJ 3  &  & $<$190 & $>$0.08  & Pleiades & MBG \cr
\+&HHJ 10 &  M5.5 & $<$300 & $>$0.08  & Pleiades  & MRM \cr
\+&HHJ 14 &   & $<$180 & $>$0.08  & Pleiades & MBG \cr
\+&BHJ 358     & M6  &  $<$300 & $>$0.065 & Hyades & MRM  \cr
\+&\cr
\hrule
\hrule
\noindent
\smallskip
References: MRM = Mart\'\i n, Rebolo, Magazz\`u (1994); MBG = Marcy, Basri,
Graham (1994); ZRMG = Zapatero-Osorio et al., this volume
\vskip 0.1truecm

\titleb{4.3} {Isolated objects in the field}

The coolest objects where Li has been searched for are several high
proper motion nearby field dwarfs (LHS~2397a, LHS~2243, LHS~2065,
LHS~2924), with spectral types in the range M8-M9 V (Kirkpatrick, Henry
\& Liebert 1993).  The mass lower limit implied by the Li
non-detections (Table 3) depends on the age assumed for these objects.
Since they are not in clusters, their age is undetermined. If they are
older than a few times $10^8$ yr their masses must be higher than
$\sim$0.065~M$_\odot$, but if any of them happens to be younger than
about $10^8$ yr its mass would have to be above the substellar mass
limit in order to account for Li depletion.
\vskip 0.1 truecm
\tabcap{3}{Li searches in field BD candidates}
\settabs\+&
Nameaaaaaaaaaaaaaaaa&aaaaaaaa&aaaaaaaaaaaaaa&aaaaaaaaaaa&aaaaaaaaaaa\cr
\+&\cr
\hrule
\hrule
\+&\cr
\+& Object & Sp.T. & LiI (m\AA) & M/M$_\odot$ & Ref.  \cr
\+&\cr
\hrule
\+&\cr
\+&LHS 1070   & M5.5 &  $<$80 &  $>$0.065 & MBG \cr
\+&LHS 36   & M6 &  $<$80 &  $>$0.065 & MBG \cr
\+&CTI 1156+28 & M7  &  $\le$400 &  $\ge$0.065 & MRM \cr
\+&VB 8  & M7 & $<$200  & $>$0.065 & MRM    \cr
\+&LHS 248   & M7 &  $<$80 &  $>$0.065 & MBG \cr
\+&TVLM 868-110639    & M7.5   & $<$500 & $>$0.065 & MRM \cr
\+&TVLM 513-46546  & M8   & $\le$300 & $\ge$0.065 & MRM \cr
\+&ESO 207-61    & M8   & $<$300 & $>$0.065 & MRM \cr
\+&LHS 2397a    & M8  &  $\le$700 & $\ge$0.065 & MRM     \cr
\+&LHS 2243    &  M8  &  $<$50 & $>$0.065 & MRM    \cr
\+&GL 569B    &  M8.5  &  $<$500 & $>$0.065 & MMR  \cr
\+&LHS 2065    &  M9  &  $<$150 & $>$0.065 & MRM   \cr
\+&LHS 2924    &  M9  &  $<$250 & $>$0.065 & MRM   \cr
\+&\cr
\hrule
\hrule
\noindent
\smallskip
References: MRM = Mart\'\i n, Rebolo, Magazz\`u (1994); MBG = Marcy, Basri,
Graham (1994); MMR =  Magazz\`u, Mart\'\i n, Rebolo (1993).
\smallskip

\titlea{5} {Concluding remarks}

Computations of  Li destruction in objects close to the substellar limit and of
the formation of Li lines using brown dwarf model atmospheres strongly support
the use of the Li I resonance line at 670.8 nm to explore the substellar nature
of brown dwarf candidates.  Any object with mass below $\sim 0.065 M\sun$ and
effective temperature higher than $\sim$ 2000 K should present a prominent Li
resonance line. At lower effective temperatures  this line is expected to be
still strong; no model atmospheres are currently available, though. The
formation of these lines in brown dwarfs spectra as well as the possible
depletion of Li into grains requires detailed investigation. Nevertheless, most
of the available brown dwarf candidates have estimated temperatures that make
suitable the observation of the Li 670.8~nm line. This feature has not been
detected in any of the searches carried out so far in these objects, which is
interpreted as  a dramatic absence of Li in all of them. The mass constraints
are more stringent for Pleiades objects than for those in binaries or in the
field. It is not ruled out that some of the very cool field objects may be a
brown dwarf, but its mass should be constrained in such case to the narrow
range 0.065-0.08 $M\sun$. At present the Li test rules out a brown dwarf nature
of the faintest proper-motion members in the Pleiades, but this cluster is very
well suited for a future detection of Li just beyond the bottom of the Main
Sequence.

\begrefchapter{References}
\ref Allard, F. (1990): Ph.D. Thesis, Univ. Heidelberg
\ref Bessell, M.S. (1991): AJ, 101, 662
\ref Boesgard, A.M., Steigman, G.  (1985): ARAA, 23, 319
\ref Bryja, C., Humphreys, R.M., Jones, T.J. (1994): AJ, 107, 246
\ref Burrows, A., Hubbard, W.B., Lunine J.I. (1989): ApJ, 345, 939
\ref Burrows, A., Hubbard, W.B., Saumon, D., Lunine, J.I. (1993): ApJ, 406, 158
\ref Burrows, A., Liebert, J. (1993): Rev. Modern Physics, 65, 301
\ref D'Antona, F., Mazzitelli, I. (1985): ApJ 296, 502
\ref Garc\'\i a L\'opez, R.J., Rebolo, R., Mart\'\i n (1994): A\&A, 282, 518
\ref D'Antona, F., Mazzitelli, I. (1994): ApJS, 90, 467
\ref Henry, T.J., McCarthy, Jr., D.W. (1993): AJ, 106, 773
\ref Ianna, P.A., Rohde, J.R.,  McCarthy, D.W. (1988): AJ, 95, 1226
\ref Jameson, R.F., Skillen, I. (1989): MNRAS, 239, 247
\ref Kirkpatrick, J.D, Henry, T.J., Liebert, J. (1993): ApJ, 406, 701
\ref Kirkpatrick, J.D, Kelly, D.M., Rieke, G.H., Liebert, J., Allard, F.,
Wehrse, R. (1993): ApJ, 402, 643
\ref Magazz\`u, A., Mart\'\i n, E.L.,  Rebolo, R. (1991): A\&A 249, 149
\ref Magazz\`u, A., Mart\'\i n, E.L., Rebolo, R. (1993): ApJ, 404, L17
\ref Magazz\`u, A.,  Rebolo, R., Pavlenko, Ya.V. (1992): ApJ, 392, 159
\ref Mart\'\i n, E.L., Magazz\`u, A., Rebolo, R. (1992): A\&A, 257, 186
\ref Mart\'\i n, E.L.,  Rebolo, R., Magazz\`u, A., Pavlenko, Ya.V. (1994a):
A\&A, 282, 503
\ref Mart\'\i n, E.L.,  Rebolo, R., Magazz\`u, A. (1994b): ApJ, Nov 20
\ref Marcy, G.W., Basri, G., Graham, J.R. (1994): ApJ, 428, L57
\ref Michaud, G. Charbonneau, P. (1991): Space Sci Rev. 57, 1
\ref Nelson, L.A., Rappaport, S., Chiang, E. (1993): ApJ, 413, 364
\ref Pavlenko, Ya.V., Rebolo, R., Mart\'\i n, E.L., Garc\'\i a L\'opez (1995):
 A\&A,  submitted
 \ref Pozio, F. (1991): Mem.\ Soc.\ Astr.\ Ital., 62, 171
\ref Rebolo, R. 1991, "Evolution of Stars: The Photospheric Abundance
Connection", eds G. Michaud and A. Tutukov, IAU Symp. 145, Kluwer, 85.
\ref Rebolo, R., Mart\'\i n, E.L., Magazz\`u, A. (1992): ApJ, 389, L83
\ref Rebolo, R., Molaro, P., Beckman, J.E. (1988) A\&A, 192, 192
\ref Simons, D.A., Becklin, E.E. (1992): ApJ, 390, 431
\ref Spite, F., Spite, M. (1982): A\&A, 115, 357
\ref Stringfellow, G.S. (1989): PhD Thesis. Univ. Calif. Santa Cruz
\ref Tinney, C.G., Mould, J.R., Reid, I.N. (1993): AJ, 105, 1045
\ref Tsuji, T. (1973): A\&A, 23, 411
\ref Wagoner, R.V., Fowler, W.A., Hoyle,F. (1967): ApJ, 148, 3

\endref

\byebye